\documentclass[aps,prl,twocolumn,superscriptaddress,reprint,showkeys]{revtex4-2}
\usepackage{graphicx}
\usepackage{amssymb, amsmath}
\usepackage{booktabs}
\usepackage[usenames]{color}
\usepackage{bm}
\usepackage{multirow}
\usepackage{dcolumn}
\usepackage{hyperref}
\usepackage{enumerate}
\usepackage[mathlines]{lineno}
\hypersetup{
    colorlinks=true,
    linkcolor=blue,
    filecolor=gray,      
    urlcolor=blue,
    citecolor=blue,
}

\begin{document}

\title{Universal materials model of deep-learning density functional theory Hamiltonian}

\newcommand{\thuphy}{State Key Laboratory of Low Dimensional Quantum Physics and Department of Physics, Tsinghua University, Beijing 100084, China}
\newcommand{\thuias}{Institute for Advanced Study, Tsinghua University, Beijing 100084, China}
\newcommand{\fscqi}{Frontier Science Center for Quantum Information, Beijing 100084, China}
\newcommand{\riken}{RIKEN Center for Emergent Matter Science (CEMS), Wako, Saitama 351-0198, Japan}
\newcommand{\buaamat}{School of Materials Science and Engineering, Beihang University, Beijing 100191, China}

\affiliation{\thuphy}
\affiliation{\thuias}
\affiliation{\fscqi}
\affiliation{\riken}
\affiliation{\buaamat}

\author{Yuxiang \surname{Wang}}
\thanks{These authors contributed equally to this work.}
\affiliation{\thuphy}

\author{Yang \surname{Li}}
\thanks{These authors contributed equally to this work.}
\affiliation{\thuphy}

\author{Zechen \surname{Tang}}
\thanks{These authors contributed equally to this work.}
\affiliation{\thuphy}

\author{He \surname{Li}}
\affiliation{\thuphy}
\affiliation{\thuias}

\author{Zilong \surname{Yuan}}
\affiliation{\thuphy}

\author{Honggeng \surname{Tao}}
\affiliation{\thuphy}

\author{Nianlong \surname{Zou}}
\affiliation{\thuphy}

\author{Ting \surname{Bao}}
\affiliation{\thuphy}

\author{Xinghao \surname{Liang}}
\affiliation{\thuphy}

\author{Zezhou \surname{Chen}}
\affiliation{\thuphy}

\author{Shanghua \surname{Xu}}
\affiliation{\thuphy}

\author{Ce \surname{Bian}}
\affiliation{\thuphy}

\author{Zhiming \surname{Xu}}
\affiliation{\thuphy}

\author{Chong \surname{Wang}}
\affiliation{\thuphy}

\author{Chen \surname{Si}}
\affiliation{\buaamat}

\author{Wenhui \surname{Duan}}
\email{duanw@tsinghua.edu.cn}
\affiliation{\thuphy}
\affiliation{\thuias}
\affiliation{\fscqi}

\author{Yong \surname{Xu}}
\email{yongxu@mail.tsinghua.edu.cn}
\affiliation{\thuphy}
\affiliation{\fscqi}
\affiliation{\riken}

\begin{abstract}
Realizing large materials models has emerged as a critical endeavor for materials research in the new era of artificial intelligence, but how to achieve this fantastic and challenging objective remains elusive. Here, we propose a feasible pathway to address this paramount pursuit by developing universal materials models of deep-learning density functional theory Hamiltonian (DeepH), enabling computational modeling of the complicated structure-property relationship of materials in general. By constructing a large materials database and substantially improving the DeepH method, we obtain a universal materials model of DeepH capable of handling diverse elemental compositions and material structures, achieving remarkable accuracy in predicting material properties. We further showcase a promising application of fine-tuning universal materials models for enhancing specific materials models. This work not only demonstrates the concept of DeepH's universal materials model but also lays the groundwork for developing large materials models, opening up significant opportunities for advancing artificial intelligence-driven materials discovery.   
\end{abstract}

\maketitle

\section{Introduction}
Density functional theory (DFT) has emerged as an exceptionally valuable first-principles approach for computational materials design, which becomes indispensable to scientific research across various disciplines, including physics, chemistry, and materials science~\cite{Martin2004}. However, due to its high computational cost, DFT calculations are typically limited to material systems of small sizes. Motivated by the Materials Genome Initiative, intensive research efforts have been dedicated to creating materials databases using DFT, while only a limited number of datasets (approximately one million solid materials) have been collected until now. The construction of extensive materials datasets and the realization of data-driven materials discovery are crucial pursuits, yet they face significant challenges at the fundamental level. 

The interplay between deep learning and DFT provides opportunities to address the aforementioned challenges~\cite{Behler2007,Zhang2018,Schutt2019,Unke2021,Gu2022,deeph2022,deeph-e32023,xdeeph2023,deeph-dfpt2024,deeph-hybrid2023,deeph22024,magnet2024,yu2023efficient}. In particular, people tried to represent the DFT Hamiltonian as a function of material structure by neural networks and train these networks with DFT computation data. By incorporating the principles of locality and equivariance into neural network design, the deep-learning DFT Hamiltonian (DeepH) approach~\cite{deeph2022} has showcased remarkable accuracy and transferability in various case studies~\cite{deeph2022,deeph-e32023,xdeeph2023,deeph-dfpt2024,deeph-hybrid2023,deeph22024,yu2023efficient,Su2023,Zhong2023}. Since all the physical quantities in the single-particle picture are derivable from the DFT Hamiltonian, the DeepH method can serve as a substitute for  computationally expensive DFT algorithms. This substitution will significantly improve computational efficiency and broaden the research scope of DFT. Moreover, unlike conventional algorithms, neural-network algorithms have the capacity to become increasingly intelligent with access to more training data. This advantage paves the way for the realization of large materials models analogous to large language models, which might revolutionize future materials discovery. Nevertheless, previous works of DeepH primarily focused on specific material systems comprising a limited number of elements. Developing a universal materials model of DeepH capable of handling most elements of the periodic table is appealing but remains elusive.

\begin{center}
\begin{figure*}
    \centering
    \includegraphics[width=\linewidth]{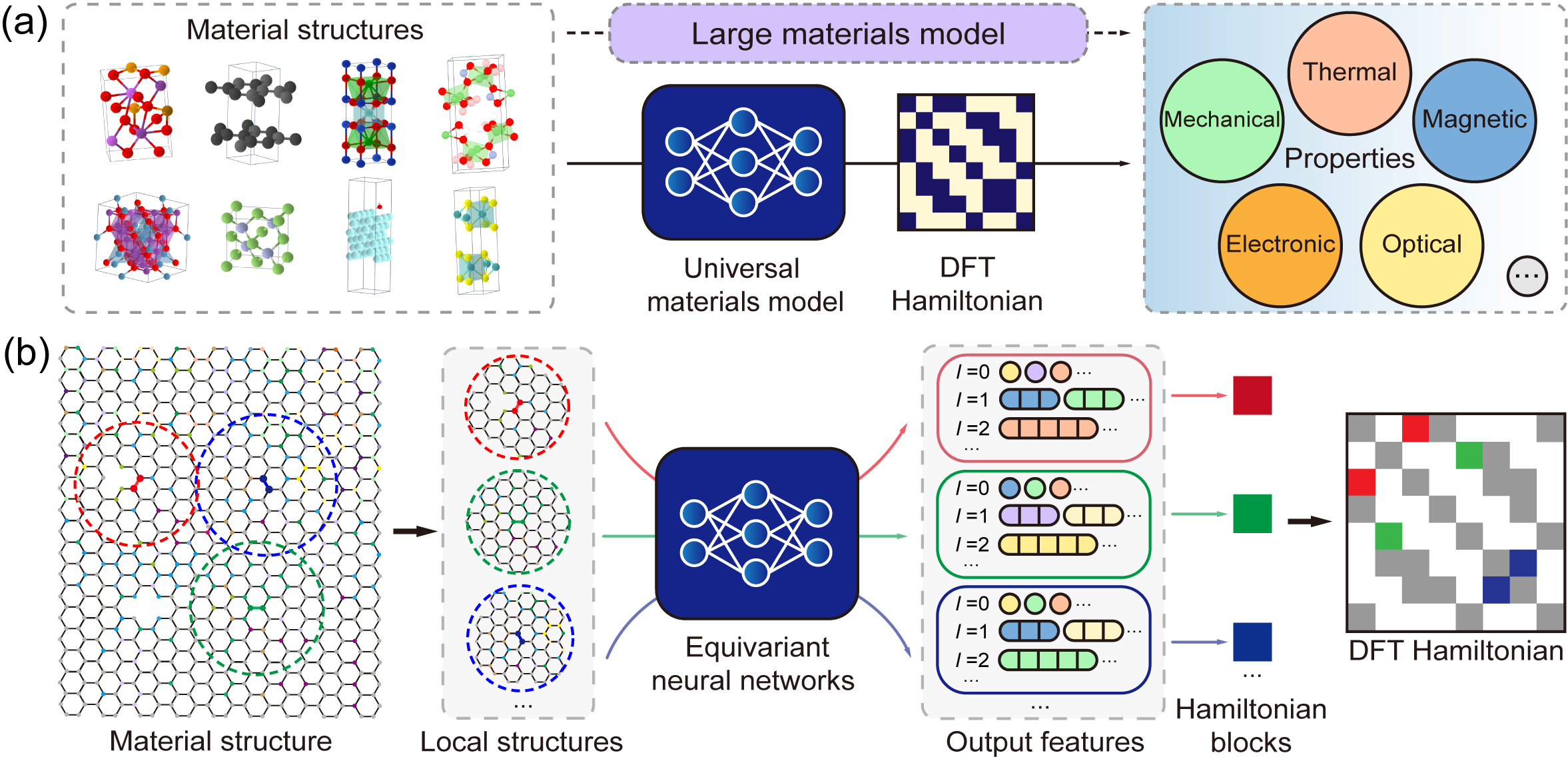}
    \caption{Sketch of universal materials model of deep-learning DFT Hamiltonian (DeepH). (a) A feasible route for developing large materials models capable of describing the structure-property relationship of materials. The universal materials model of DeepH accepts an arbitrary material structure as input and generates the corresponding DFT Hamiltonian, enabling straightforward derivation of various material properties. (b) Working principle of DeepH, which learns and predicts DFT Hamiltonian matrix blocks separately based on local-structure information. In equivariant neural networks, output features labelled by different angular quantum number $l$ are utilized to represent the DFT Hamiltonian.}
    \label{fig1}
\end{figure*} 
\end{center}

In this work, we employed the automated interactive infrastructure and database (AiiDA) approach to construct DFT training datasets comprising $\sim$$10^4$ solid materials~\cite{AiiDA2020,AiiDA2021}, with elements spanning the first four rows of the periodic table. Subsequently, we employed the DeepH-2 method~\cite{deeph22024}, an advanced equivariant transformer architecture that harnesses attention mechanisms derived from local-coordinate system information, to train a neural network model of DeepH using the DFT datasets. Furthermore, we introduced the gauge equivalence of the DFT Hamiltonian into neural networks and adjusted the neural network's loss function, which helps address the complexities of training tasks relevant to diverse elemental compositions and varying material structures. All these endeavors culminate in the training of a universal materials model of DeepH, yielding a notably low averaged mean absolute error (MAE) of 2.2~meV, comparable to those obtained for specific materials models. Importantly, our results demonstrate a strong negative correlation between the number of training structures and the MAE, indicating that model accuracy can be improved by utilizing larger training datasets. As an illustrative application, we fine-tuned the universal materials model to handle a large dataset of carbon materials, resulting in a highly accurate DeepH model with limited training data. 
The work showcases the feasibility of training a universal materials model of DeepH, opening up opportunities for developing large materials models as well as for advancing artificial intelligence (AI)-driven materials discovery.

Following the success of large language models, the concept of large materials models as deep-learning computational models for materials design has attracted great interests. Nevertheless, the task of acquiring large materials models appears to be quite challenging, given the inherent complexity of the structure-property relationship in materials. Materials databases are demanded for developing large materials models, which could be constructed through experiments or first-principles calculations. The latter way is generally more controllable and cost-effective, making it preferable for this purpose. Among first-principles methods, DFT is unequivocally the preferred choice due to its balance of accuracy and efficiency. Therefore, several well-known materials databases have been constructed using DFT under the Materials Genome Initiative, including the Materials Project~\cite{MaterialsProject}, NOMAD~\cite{NOMAD}, AFLOW~\cite{AFLOW}, and others. In this context, the development of large materials models based on DFT and state-of-the-art deep learning methods becomes imperative.

A critical issue is to identify a viable pathway for integrating deep learning and DFT methods to develop large materials models. A general strategy is to select material structures as inputs and DFT calculation results as outputs, and utilize neural networks to model the input-output relationship. By employing DFT calculation results of diverse material structures for training, properties of new material structures can be predicted based on the trained networks. Since several types of physical quantities are available from DFT, a natural question arises: Which specific physical quantities in DFT are preferred for developing large materials models?

In our opinion, the DFT Hamiltonian is a desired target quantity for deep-learning DFT (Fig.~\ref{fig1}), as will be elaborated. Firstly, the DFT Hamiltonian is a fundamental quantity from which all other physical quantities of DFT can be derived straightforwardly. These include the total energy, charge density, band structure, physical responses, and more~\cite{Martin2004}, as illustrated in Fig.~\ref{fig1}a. Secondly, owing to the localized nature of the DFT Hamiltonian~\cite{Kohn1996,Prodan2005}, the problem of deep-learning DFT can be considerably simplified. Specifically, the DFT Hamiltonian under the localized atomic-like basis can be represented as a sparse matrix, with its matrix elements determined by local chemical environments. Thus, one can model the Hamiltonian matrix elements between neighboring pairs of atoms based on their nearby structural information, rather than modeling the entire DFT Hamiltonian matrix for the entire material structure, as illustrated in Fig.~\ref{fig1}b. This not only significantly simplifies the deep learning task but also greatly enhances the amount of training data, as one material structure contains numerous nonzero Hamiltonian matrix elements. On the inference side, the trained model can generalize well to new material structures unseen in the training data once the networks have learned enough training data for varying local chemical environments. The essential physics is summarized by the locality principle, or the ``nearsightedness'' principle explicitly proposed by Walter Kohn~\cite{Kohn1996,Prodan2005}, which is the key to developing highly transferable deep-learning models.

Models of DeepH have been developed for specific material systems, typically comprising less than four elements and including atomic structures closely related to a few mother materials~\cite{deeph2022,deeph-e32023,xdeeph2023,deeph-dfpt2024,deeph-hybrid2023,deeph22024}. We refer to such models as specific materials models of DeepH. Differently, universal materials models of DeepH correspond to general material systems containing diverse elemental compositions and atomic structures. With such models, the structure-property relationship of materials in general can be described within the framework of deep-learning DFT, which paves the way for realizing large materials models. This motivates us to develop a universal materials model of DeepH.

Compared to specific materials models of DeepH, developing universal materials models is challenged by the following facts: (i) Although large materials databases computed by DFT have been constructed before, data of DFT Hamiltonians are often not saved or inaccessible. Thus, the required large amount of training data must be generated from scratch. (ii) The training task involving over tens of elements and varying material structures has rarely been attempted before. It remains uncertain whether the current neural network architectures are competent for this task. (iii) The DFT Hamiltonian $\hat{H}_{\text{DFT}}$ is subjected to a gauge freedom such that a change of energy reference $\hat{H}_{\text{DFT}} \to \hat{H}_{\text{DFT}} +\Delta \mu$ does not affect physics, where $\Delta \mu$ is a constant number~\cite{Martin2004}. The DeepH approach applies neural networks to represent the mapping from material structure $\{\mathcal R\}$ to $\hat{H}_{\text{DFT}}$. The gauge typically varies among DFT calculations of different material structures. The fundamental ``gauge problem'' becomes critical when handling diverse material structures, which should be addressed for realizing a universal materials model of DeepH.  In the forthcoming sections, we will endeavor to surmount the challenges mentioned above.

\section{Methods}

\begin{center}
\begin{figure*}
    \centering
    \includegraphics[width=\linewidth]{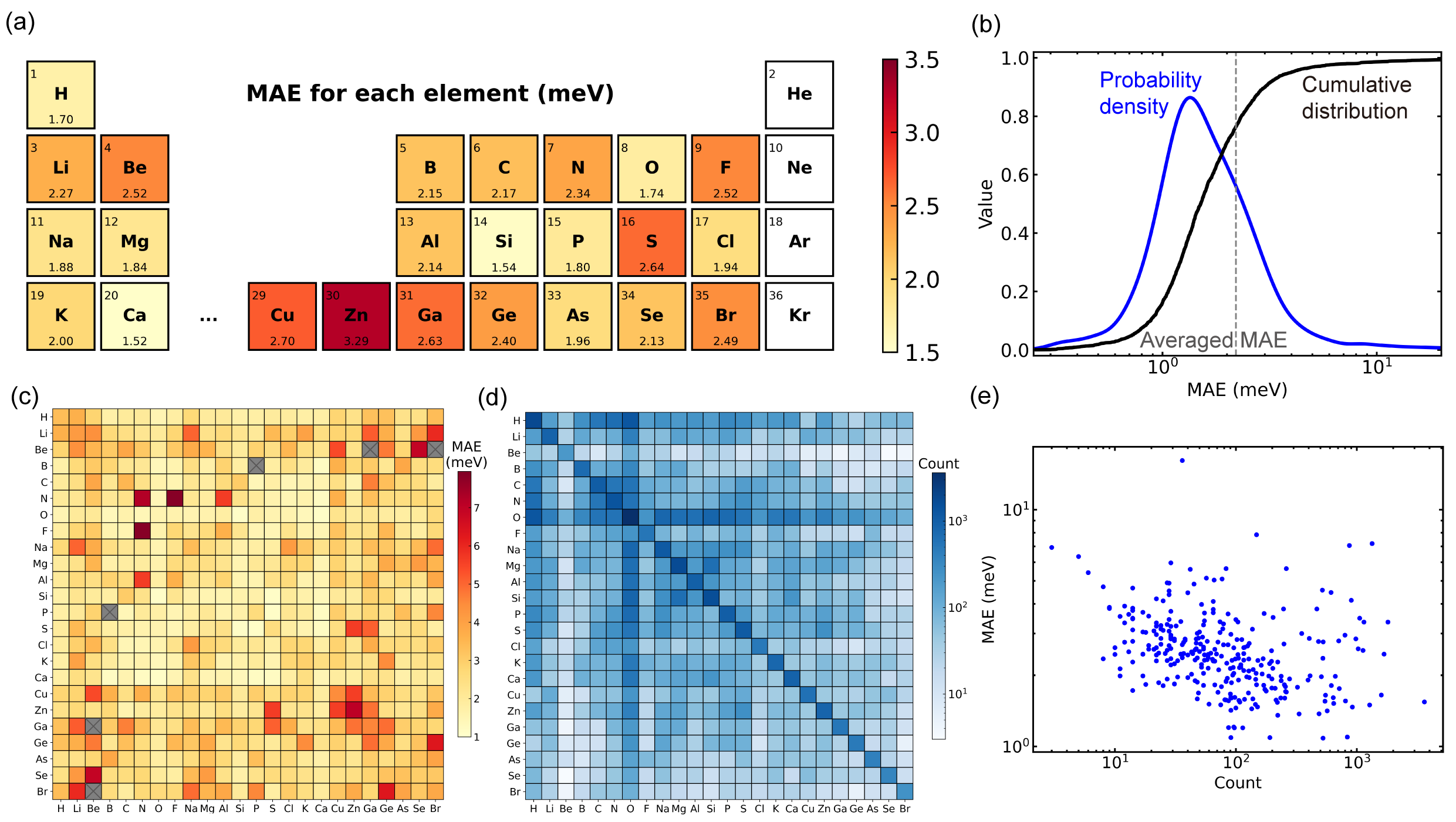}
    \caption{Performance evaluation of the universal materials model trained by the DeepH-2 method using a large materials database comprising $\sim$$10^4$ solid materials. (a) MAEs of the predicted DFT Hamiltonian averaged for each element, showing high prediction accuracy at the level of meV. (b) Cumulative distribution and probability density of MAEs across test structures, giving an averaged MAE of 2.2~meV. (c,d) Distribution of (c) MAEs and (d) counts of training structures accross element pairs. In (c), the element pairs corresponding to seldom or zero test structures are marked in gray. (e) Scatter plot depicting the relationship between counts of training structures and MAEs of element pairs, revealing a strong negative correlation as indicated by Spearman's rank analysis.}
    \label{fig2}
\end{figure*} 
\end{center}

To illustrate the concept of universal materials model of DeepH, we first prepare a large materials database comprising $\sim$$10^4$ solid materials. As the first attempt, we simplify the study by neglecting the spin-orbit coupling and excluding magnetic materials. A further generalization of DeepH to take these issues into account is straightforward~\cite{deeph-e32023,xdeeph2023}. To showcase a diverse elemental composition, we choose the first four rows of the periodic table as displayed in Fig.~\ref{fig2}a, excluding transition elements from Sc to Ni to avoid magnetism and excluding noble gas elements. Candidate material structures are sourced from the database of Materials Project~\cite{MaterialsProject}. In addition to filtering based on element types, the candidates are further refined to include only those labeled ``non-magnetic" in the Materials Project. Structures including over 150 atoms in the unit cell are excluded for simplicity. As a result of these filtering criteria, the final materials dataset consists of a total of 12,062 structures, which will be considered in this work. Statistics for the dataset structures, including distribution of atom number and element number is displayed in supplementary Figs.~S2 and S3 (online).

Next, we develop a high-throughput workflow for DFT calculations utilizing the  framework of AiiDA (Automated Interactive Infrastructure and Database) ~\cite{AiiDA2020,AiiDA2021}, and employ it to build the materials database. The DFT calculations are performed by the OpenMX package~\cite{openmx2003}, using norm-conserving pseudopotentials and pseudo-atomic bases. High-throughput parameters for DFT calculations are selected to ensure a high level of accuracy and data quality, with method details described in the supplementary material.

The DeepH approach~\cite{deeph2022} has been developed and extensively studied~\cite{deeph-e32023,xdeeph2023,deeph-dfpt2024,deeph-hybrid2023,deeph22024,magnet2024,Schutt2019,Unke2021,Gu2022,yu2023efficient}. Recently, the method has been generalized for calculations using beyond DFT methods~\cite{deeph-hybrid2023} and density functional perturbation theory~\cite{deeph-dfpt2024} as well as for studying magnetic materials~\cite{xdeeph2023,magnet2024,deltaspin2023}. In DeepH, the key idea is to employ neural networks to represent $\hat{H}_{\text{DFT}}(\{\mathcal R\})$ as a function of material structure $\{\mathcal R\}$. By varying the input material structure, training data of  $\hat{H}_{\text{DFT}}$ are initially generated by DFT codes and subsequently utilized to train neural networks. These trained network models are then employed to make inferences on new material structures. Some prior knowledge is critical to DeepH and should be properly considered, as will be discussed.

A crucial prior knowledge is the locality (or ``nearsightedness'') principle, as discussed earlier. To leverage the locality principle, we express the DFT Hamiltonian under the localized atomic-like basis, and  decompose the Hamiltonian into blocks that describe inter- or intra-atomic coupling. Thus, a single training material structure may correspond to a substantial amount of data for Hamiltonian blocks. In addition, each Hamiltonian block can be determined based on the information of local structures rather than the entire structure, as illustrated in Fig.~\ref{fig1}b. The simplification greatly facilitates deep learning, ensuring the high accuracy and transferability of DeepH models.

Another fundamental prior knowledge is the equivariance (or symmetry) principle, which asserts that the laws of physics remain unchanged when observed from different coordinate systems. Consequently, the corresponding physical quantities and equations exhibit equivariance under coordinate transformations. Here, the relevant transformation group is the Euclidean group in three-dimensional space $\text{E}(3)$, which includes spatial translation, inversion, and rotation. Both the input material structure and the output Hamiltonian matrix under the localized atomic-like basis change equivariantly under coordinate transformations. Preserving the equivariance property not only improves data efficiency but also enhances generalization ability, which can significantly enhance the performance of DeepH. The first generation architecture of DeepH simplifies the equivariant problem into an invariant one by utilizing local coordinate systems and restores the equivariant feature through local-coordinate transformations~\cite{deeph2022}. The next generation architecture of DeepH is based on equivariant neural networks, named the DeepH-E3 method~\cite{deeph-e32023}. In this framework, all the feature vectors of input, hidden, and output layers are equivariant vectors. The interactions between feature vectors are described by equivariant tensor products. The DFT Hamiltonian matrix is formed from output feature vectors using the Wigner–Eckart theorem.

Recently, a new-generation architecture of DeepH, named DeepH-2, has been proposed by some authors of this work~\cite{deeph22024}. DeepH-2 shows better performance than its predecessors in terms of both efficiency and accuracy. Firstly, local coordinate systems are introduced to enhance efficiency. Unlike the first generation architecture of DeepH, 
in DeepH-2, only the local $z$-axis is selected for each target edge to align with the direction of the associated atom pairs, while the remaining two axes are left undetermined. The general challenge of defining local coordinate systems for arbitrary local structures can thus be overcome. Secondly, the adoption of local coordinate systems leads to the reduction of the three-dimensional rotation group SO(3) to the two-dimensional rotation group SO(2), which considerably simplifies the implementation of rotation equivariance in neural networks, as previously proposed~\cite{escn2023}. 
This efficient tensor product scheme has been applied in combination with the transformer architecture to develop neural network force fields~\cite{liao2024equiformerv2}. Thirdly, the cutting-edge equivariant transformer architecture is integrated into DeepH-2, where attention mechanisms are computed using the local-coordinate information. These enhancements of DeepH-2 yield state-of-the-art performance, as evidenced by experiments conducted on specific material systems~\cite{deeph22024}. We will apply the DeepH-2 approach to train the universal materials model.

A prior knowledge of gauge freedom in physical quantities is often overlooked, yet it plays a fundamentally important role in deep learning. Here, the DFT Hamiltonian possesses gauge freedom, meaning that two DFT Hamiltonians are considered equivalent if they differ by a constant $\Delta \mu$. Only by acknowledging this gauge equivalence can the mapping from material structure to the DFT Hamiltonian be described as a function. However, previous works usually used a loss function of the form MSE($H^{\theta}_{\text{DFT}}$, $H^{0}_{\text{DFT}}$), which denotes the mean squared error (MSE) between the predicted DFT Hamiltonian matrix $H^{\theta}_{\text{DFT}}$ and the labelled one $H^{0}_{\text{DFT}}$. This loss function is gauge dependent, which could potentially lead to a ``gauge problem'' arising from the inherent gauge arbitrariness in DFT Hamiltonians. While the gauge problem may not seem serious when modeling specific materials datasets with minor structural perturbations, it becomes critical when developing a universal materials model of DeepH.

To tackle this critical ``gauge problem", the loss function should be appropriately defined to incorporate gauge equivalence. In fact, the metric that measures the distance between two DFT Hamiltonian matrices should be redefined. For instance, if two DFT Hamiltonians are equivalent, their distance should be zero. Thus, a new form of loss function MSE($H^{\theta}_{\text{DFT}}$, $H^{0}_{\text{DFT}} + \Delta \mu * S$) is introduced, where $S$ denotes the overlap matrix. Let $H_{\alpha\beta}$ and $S_{\alpha\beta}$ denote the elements of DFT Hamiltonian matrix and overlap matrix under the localized atomic-like basis, respectively. In our calculations, $\Delta \mu$ is determined via the formula: $\Delta \mu =\frac{\text{Re}\left(\sum_{\alpha\beta}(H_{\alpha\beta}^{\theta}-H_{\alpha\beta}^{0})S^*_{\alpha\beta}\right)}{\sum_{\alpha\beta}S_{\alpha\beta}S^*_{\alpha\beta}}$, where $\text{Re}$ denotes the real part~\cite{deeph22024}. With this strategy, the loss function becomes gauge-invariant, effectively addressing the ``gauge problem". The displayed MAE in this work has already taken this gauge correction into account, defined as MAE($H^\theta,H^0+\Delta\mu * S$).

Our universal materials model of DeepH is trained by the DeepH-2 approach~\cite{deeph22024} using 17.28 million parameters. The neural networks are composed of $3$ equivariant transformer blocks for message passing, with each node and edge carrying $80$ equivariant features up to $l=4$. $l$ denotes the angular quantum number. The number of attention heads for transformer blocks is $2$. The embedding of material structure includes embeddings of atomic numbers and inter-atomic distances  with strategy of Gaussian smearing of 600 bases with center ranging from 0.0 to 9.0~\text{\AA}. The output features of neural networks are passed through a linear layer and then employed to construct DFT Hamiltonians through a ``Wigner-Eckarts" layer~\cite{deeph-e32023}. The training process was carried out on an NVIDIA A100 GPU and stopped at 343 epochs, taking 207 hours. Apart from utilizing the ``gauge-invariant" loss introduced above, the Huber loss with a Huber delta of 0.01~eV is applied~\cite{huber1964}. Throughout the training process, the batch size is fixed at 1, meaning that each batch contains one material structure. An AdamW optimizer with betas of $(0.9, 0.999)$ and weight decay of $1\times 10^{-3}$ is applied, with ``ReduceLROnPlateau" scheduler. The initial learning rate is $4\times 10^{-4}$, with a decay rate of 0.5 and decay patience 20. The minimal learning rate is selected to be $1\times 10^{-5}$, and the training terminates when the learning rate reaches this value.

\section{Results}

Until now, we have prepared a large materials database comprising $\sim$$10^4$ solid materials, encompassing diverse elemental compositions and material structures. Utilizing the materials database, we train a universal materials model of DeepH using the cutting-edge DeepH-2 method and adapting the loss function to address the ``gauge problem''. The dataset is split into training, validation, and test structures with a ratio of $6:2:2$ in the course of training. All displayed prediction results are selected from test sets, unseen in the training process.

\begin{center}
\begin{figure*}
    \centering
    \includegraphics[width=\linewidth]{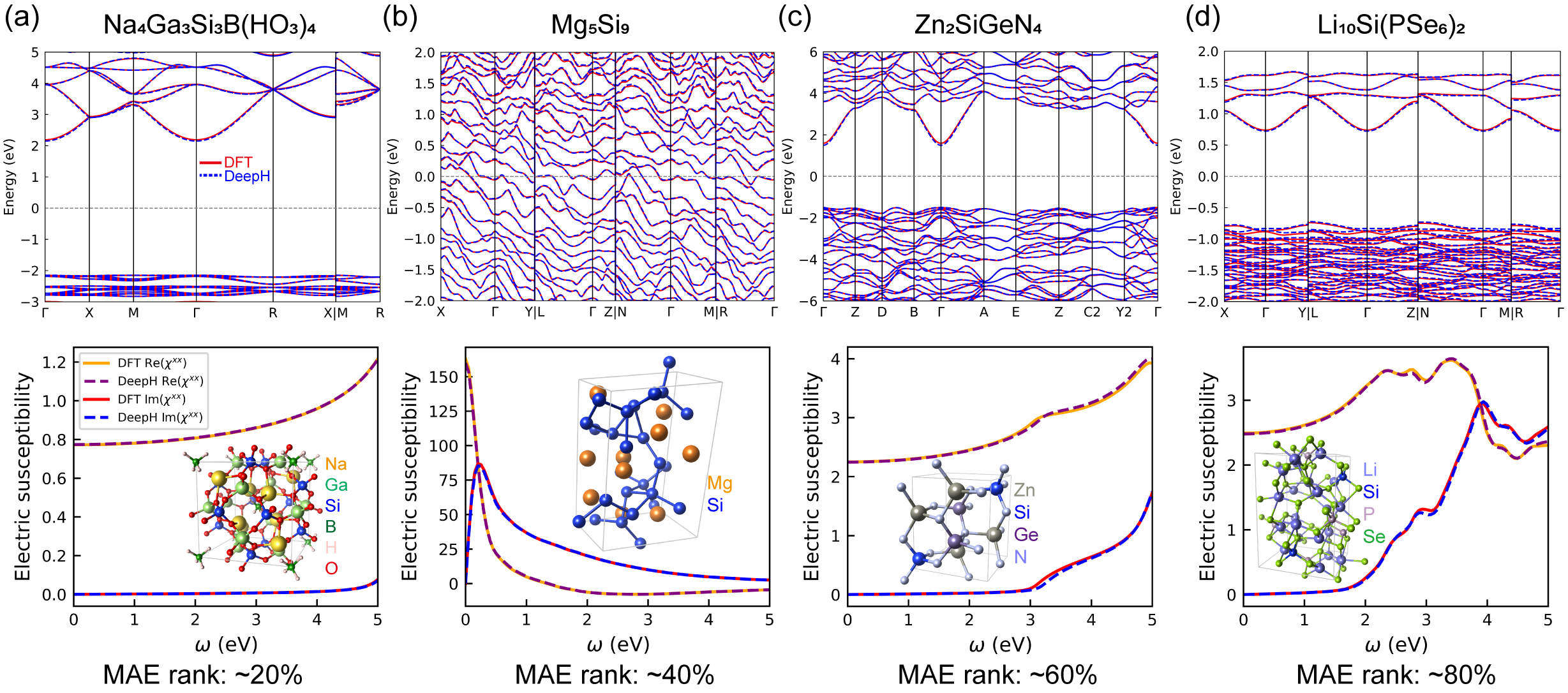}
    \caption{Testing the universal materials model by comparing DFT-calculated and DeepH-predicted band structures and electric susceptibilities $\chi$ as a function of frequency $\omega$ for representative test materials (with MAE ranks of $\sim20\%$, $\sim40\%$, $\sim60\%$ and $\sim80\%$). (a-d) Comparison results for (a) Na$_4$Ga$_3$Si$_3$B(HO$_3$)$_4$ (Materials Project ID: mp-534870), (b) Mg$_5$Si$_9$ (mp-1075644), (c) Zn$_2$SiGeN$_4$ (mp-1215642), (d) Li$_{10}$Si(PSe$_6$)$_2$ (mp-721253). The corresponding material structures are displayed in the insets.}
    \label{fig3}
\end{figure*} 
\end{center}

Let us evaluate the performance of the trained universal materials model of DeepH. Across the training, validation, and test sets, MAEs of the predicted DFT Hamiltonian matrix elements reach 1.45, 2.35, and 2.20~meV, respectively. Only a marginal degree of overfitting is observed, indicating that the model has capacity to generalize to unseen structures. Remarkably, the averaged MAE of 2.2~meV is as low as those for specific materials models trained by the first generation architecture of DeepH~\cite{deeph2022}. This is quite impressive, especially considering that the deep-learning task of the universal materials model is significantly more challenging.

In Fig.~\ref{fig2}a and b, a breakdown of the element-wise and structure-wise MAEs of the predicted Hamiltonian matrix elements is presented. The element-wise MAE represents the MAE averaged across material structures containing a specific element, which ranges from $1.52$ to $3.29$~meV, indicating consistent performance across all elements. The structure-wise MAE distribution in Fig.~\ref{fig2}b reveals a biased distribution across material structures. Specifically, approximately $80\%$ of material structures have MAEs smaller than the averaged MAE (2.2~meV) across the dataset. Only 34 structures (about $1.4\%$ of the test set) have MAEs exceeding $10$~meV, indicating the model's good prediction accuracy on mainstream structures.

Further analysis of the dataset indicates that the bias in the model's performance on material structures may be attributed to the bias in the dataset distribution. Figure~\ref{fig2}c and d displays the distributions of MAEs of the predicted DFT Hamiltonian and counts of training structures across element pairs, respectively. A general trend is observed: The more training structures of element pairs included in the dataset, the smaller the corresponding MAEs. The relationship between these two quantities is depicted in Fig.~\ref{fig2}e. By performing Spearman's rank analysis on the data, we obtain a Spearman's correlation coefficient of -0.3692 and a $p$-value of $4.041\times10^{-11}$, revealing a strong negative correlation between MAEs and counts of training structures. This phenomenon may suggest the existence of a ``scaling rule" for universal materials models of DeepH, indicating that larger training datasets could lead to better model performance.

To assess the accuracy of our universal materials model of DeepH in predicting material properties, we employ both DFT-calculated and DeepH-predicted DFT Hamiltonians for computing example physical quantities, including band structures and electric susceptibilities as a function of frequency, with more details described in supplementary material. We then compare the calculated results obtained from both methods. A few representative material structures are selected from the test set for this comparison study. These material structures are chosen based on their ranking in terms of MAE. Their MAE ranks are about 20$\%$, 40$\%$, 60$\%$, and 80$\%$, with a margin of $\pm{}1{}\%$. Moreover, the selected material structures feature complex chemical compositions and atomic structures, presenting a challenging test for DeepH. As shown in Fig.~\ref{fig3}, the DeepH-predicted results closely match the DFT-calculated ones for the example material studies, showcasing the impressive predictive accuracy of DeepH in calculating material properties.

\begin{center}
\begin{figure*}
    \centering
    \includegraphics[width=0.8\linewidth]{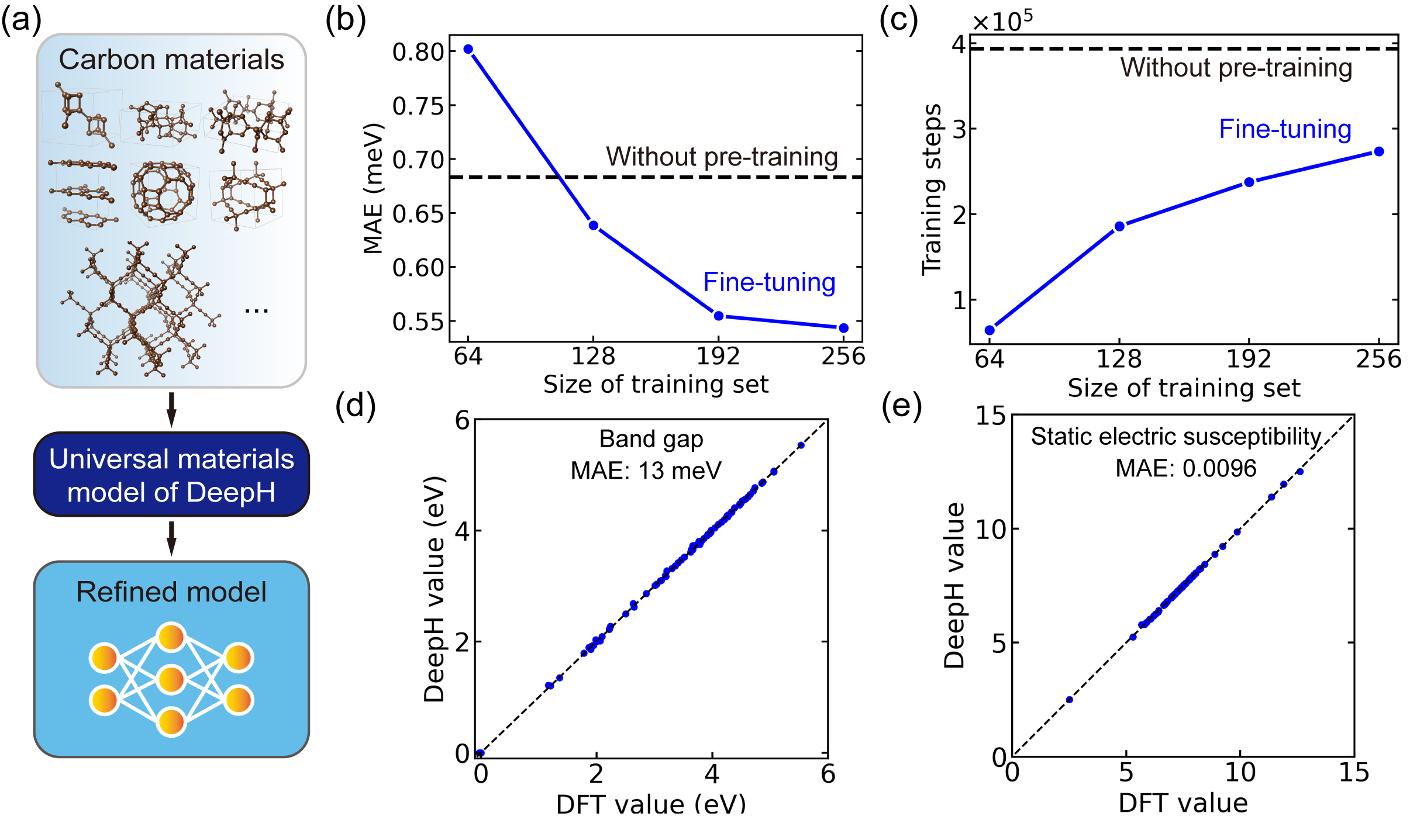}
    \caption{Fine-tuning the universal materials model for studying specific materials datasets. (a) Schematic workflow of fine-tuning. A specified materials dataset containing carbon allotropes is supplied to the universal materials model of DeepH, and a refined model is yielded after fine-tuning. (b,c) Relationship between (b) test-set MAE or (c) training steps and the size of fine-tuning training set, compared with a model obtained without pre-training (black dased line). (d,e) Comparisons of (d) band gap and (e) static electric susceptibility calculated by DFT and predicted by DeepH for the test set of carbon allotropes. DeepH shows low MAEs of 13~meV and 0.0096, respectively.}
    \label{fig4}
\end{figure*} 
\end{center}

Despite the overall success shown by our universal materials model of DeepH, notable limitations are evident in the preliminary results. For instance, the material structure with the highest MAE has an MAE of 130.6 meV in the predicted DFT Hamiltonian, and there exist another two test structures showing MAE values exceeding 100~meV. This problem could be mitigated by using a refined test set as discussed in in supplementary material. Moreover, it is still inevitable to encounter test material structures with inaccurately predicted band structures, due to the large MAEs in the predicted DFT Hamiltonians and/or the singularity in the overlap matrix. Further improvements on the universal materials model are feasible, for instance, by increasing the amount of training data. Furthermore, fine-tuning the universal materials model provides a promising way to enhance the model's performance in studying specific materials databases, as we will demonstrate.

As a case application, we fine-tune the universal materials model to study carbon allotropes. The carbon materials dataset are sourced from the Samara Carbon Allotrope Database (SACADA)~\cite{SACADA2016}, which consists of 427 carbon allotropes with diverse atomic structures as illustrated in Fig.~\ref{fig4}a. The example study on the SACADA database can be viewed as a case study on material structures with defects, as it includes various kinds of carbon allotropes with defects in the dataset.
We supply varying amounts of training data on carbon allotropes to the universal materials model trained previously and fine-tune it to create a refined DeepH model specifically tailored for carbon materials (Fig.~\ref{fig4}a), with neural network hyperparameters described in supplementary material. The carbon materials dataset is divided into training, validation and test sets with a ratio of $6:2:2$. For comparison, four fine-tuned DeepH models are obtained using $25\%$, $50\%$, $75\%$, and $100\%$ of the training data, and one DeepH model is trained from scratch using the entire training dataset. All these models are evaluated on the same test set comprising 85 carbon allotropes, none of which have been seen during the training or fine-tuning processes. Despite the diverse atomic structures included in the SACADA database, the universal materials model of DeepH exhibits a low zero-shot generalization error of 1.8~meV, demonstrating its good transferability.

The performance of the fine-tuned models is depicted in Fig.~\ref{fig4}b and c. Compared to the model without pre-training, fine-tuning can significantly reduce the MAE in the predicted DFT Hamiltonian to as low as 0.54~meV; it can also achieve comparable prediction accuracy by using less than $50\%$ of the training structures (Fig.~\ref{fig4}b). Orbital-averaged MAE is also lowered, as compared in supplementary Fig.~S4 (online). Additionally, fine-tuning also leads to significant improvements in training convergence and reduces training time, as visualized in Fig.~\ref{fig4}c, showing a $\sim$30\% saving of training steps even for fine-tuning with the full training dataset. In this sense, fine-tuning is helpful for improving prediction accuracy and enhancing training efficiency.

Importantly, the DeepH model after fine-tuning performs remarkably well in predicting material properties. Here, we design a high-throughput workflow to calculate the band gaps and the traces of the static electric susceptibility tensor for all the test structures using both the DFT-calculated and DeepH-predicted DFT Hamiltonians, as described in the supplementary material. Comparisons of the results obtained from the two methods are displayed in Fig.~\ref{fig4}d and e. Note that the static electric susceptibility diverges for materials with gapless band structures and those results are excluded in Fig.~\ref{fig4}e. Despite the comparatively diverse distribution of the result data, the fine-tuned DeepH model can predict band gaps and static electric susceptibilities with very low MAEs of 13~meV and 0.0096, respectively. Furthermore, we observe that fine-tuning can help address inaccuracies in predicting band structures (see supplementary Fig.~S5 (online)), and the fine-tuned model can provide accurate predictions of band structures for almost all the test structures (see supplementary Fig.~S6 (online)). With these significant improvements, we believe that the universal materials model in combination with fine-tuning will find important applications in future materials research.

\section{Discussion and conclusion}
In conclusion, we have constructed a large materials database, applied state-of-the-art neural network architecture and made significant improvements on the DeepH approach. Through these efforts, we demonstrate the feasibility of training a universal materials model of DeepH, achieving high accuracy in predicting DFT Hamiltonians and material properties for most test materials. Furthermore, as a promising application, we showcase that fine-tuning the universal materials model can find useful applications in specific materials research.

As a future perspective, let us discuss a few important directions for further research. Firstly, it is expected that the ``scaling law'' of achieving better model performance with larger datasets and models will take effect at the current stage of materials research. Thus, a critical task is to continuously expand the size of universal materials models of DeepH. Secondly, a notable limitation of the current model is the neglect of spin-orbit coupling and the exclusion of magnetic structures. These important issues could potentially be addressed by universal materials models of DeepH~\cite{deeph-e32023,xdeeph2023} for covering the whole periodic table, which awaits future works. Last but not least, DeepH offers an efficient way to construct large materials databases containing comprehensive physical properties, such as Bardeen–Cooper–Schrieffer superconductivity~\cite{deeph-dfpt2024}, which could find wide applications in future materials research. For instance, it could be utilized for searching superconducting materials that operate at high temperatures.

Finally, we would like to emphasize the significance of this work. The DeepH approach enables the substitution of DFT algorithms by neural network models, offering the potential to significantly enhance computational efficiency by orders of magnitude~\cite{deeph2022,deeph-e32023,deeph-hybrid2023,deeph-dfpt2024}. Furthermore, as it gains access to larger materials databases, it can also become increasingly intelligent. The latter ability is of critical importance to materials discovery but has only been demonstrated for specific material models before. It is this work that showcases the feasibility of training a universal materials model across diverse elemental compositions and material structures. This will be fundamentally important because DeepH can not only accurately model the mapping function from material structure to DFT Hamiltonian but also be used to efficiently derive material properties. Therefore, the complicated structure-property relationship of materials in principle could be described by DeepH. In this sense, the realization of universal materials models of DeepH paves the way for the development of large materials models, offering great opportunities for AI-driven materials discovery.  

\emph{Note: During the preparation of this work, we became aware of a related study~\cite{zhong2024universal}.}

\section{Conflict of interest}
The authors declare no conflict of interest.

\section{Acknowledgments}
This work was supported by the Basic Science Center Project of NSFC (grant no. 52388201), the National Natural Science Foundation of China (grant no. 12334003), the National Science Fund for Distinguished Young Scholars (grant no. 12025405), the Ministry of Science and Technology of China (grant no. 2023YFA1406400), the Beijing Advanced Innovation Center for Future Chip (ICFC), and the Beijing Advanced Innovation Center for Materials Genome Engineering. Yang Li is funded by the Shuimu Tsinghua Scholar program. The work was carried out at National Supercomputer Center in Tianjin using the Tianhe new generation supercomputer.

\section{Author contributions}
Yong Xu and Wenhui Duan proposed the project and supervised Yuxiang Wang, Yang Li and Zechen Tang in carrying out the research, with the help of all other co-authors. All authors discussed the results and participated in the preparation of the manuscript.

\end{document}